\documentclass[singlespacing]{elsart}
\usepackage{epsfig}

\begin{document}

\begin{frontmatter}

\title{Development of CaMoO$_4$ crystal scintillators for double beta decay experiment with $^{100}$Mo}

\author[ISA] {A.N.~Annenkov},
\author[ISA] {O.A.~Buzanov},
\author[INR-Kiev] {F.A.~Danevich\thanksref{1}}
 \thanks[1]{Corresponding author.
 Address: Institute for Nuclear Research, Prospect Nauky 47, MSP 03680 Kyiv, Ukraine;
 tel: +380-44-525-1111;
 fax: +380-44-525-4463;
 E-mail address: danevich@kinr.kiev.ua},
 \author[INR-Kiev] {A.Sh.~Georgadze},
\author[DMRC] {S.K.~Kim},
\author[Daegu] {H.J.~Kim},
\author[Sejong] {Y.D.~Kim},
\author[INR-Kiev] {V.V.~Kobychev},
\author[ITEP] {V.N.~Kornoukhov},
\author[INP-Minsk]{M.~Korzhik},
\author[Sejong] {J.I.~Lee},
\author[INP-Minsk]{O.~Missevitch},
\author[INR-Kiev] {V.M.~Mokina},
\author[INR-Kiev] {S.S.~Nagorny},
\author[INR-Kiev] {A.S.~Nikolaiko},
\author[INR-Kiev] {D.V.~Poda},
\author[INR-Kiev] {R.B.~Podviyanuk},
\author[INR-Kiev] {D.J.~Sedlak},
\author[INR-Kiev] {O.G.~Shkulkova},
\author[Daegu] {J.H.~So},
\author[IM] {I.M.~Solsky},
\author[INR-Kiev] {V.I.~Tretyak},
\author[INR-Kiev] {S.S.~Yurchenko}

\address[ISA] {Moscow Steel and Alloy Institute, 119049, Moscow, Russia}
\address[INR-Kiev]{Institute for Nuclear Research, MSP 03680 Kyiv, Ukraine}
\address[DMRC] {DMRC and School of Physics, Seoul National University,
Seoul, 151-742, Republic of Korea}
\address[Daegu] {Physics Department, Kyungpook National University, Daegu, 702-701, Republic of Korea}
\address[Sejong] {Sejong University, Seoul, Republic of Korea}
\address[ITEP] {Institute for Theoretical and Experimental Physics, 117218 Moscow, Russia}
\address[INP-Minsk] {Institute for Nuclear Problems, 220030 Minsk, Belarus}
\address[IM] {Institute for Materials, 79031 Lviv, Ukraine}

\begin{abstract}

Energy resolution, $\alpha/\beta$ ratio, pulse-shape
discrimination for $\gamma$ rays and $\alpha$ particles,
temperature dependence of scintillation properties, and
radioactive contamination were studied with CaMoO$_4$ crystal
scintillators. A high sensitivity experiment to search for
$0\nu2\beta$ decay of $^{100}$Mo by using CaMoO$_4$ scintillators
is discussed.

\end{abstract}

\end{frontmatter}

\section{INTRODUCTION}

As it was already demonstrated by several experiments,
scintillation method is a promising tool to search for the double
beta (2$\beta$) decay processes
\cite{Dan89,Bur95,Dan96,Bel99,GSO,Dan03,Bel03,Oga04,ZWO}.
Scintillation detectors possess a range of important
characteristics for a high sensitivity 2$\beta$ decay experiment:
high registration efficiency for $2\beta$ processes, reasonable
energy resolution, pulse-shape discrimination ability to reduce
background, operating stability, low cost. There exists a few
detectors containing Molybdenum. The most promising of them, from
the point of view of light output, is calcium molybdate
(CaMoO$_4$). In Ref. \cite{Belo05} CaMoO$_4$ crystal scintillators
were proposed to search for neutrinoless ($0\nu$) double beta
decay of $^{100}$Mo. Recently CaMoO$_4$ was intensively studied as
possible cryogenic scintillation-bolometric detector for
experiments to search for $2\beta$ decay and dark matter
\cite{Mikh05-pssb,Mikh05-JAP,Mikh05-JPCM,Seny06,Mikh06-JPDAP,Pirr06}.

$^{100}$Mo is one of the most promising candidate for $2\beta$
decay experiments because of its high transition energy
($Q_{2\beta}$=3035 keV). As a result, the calculated value of the
phase space integral $G_{mm}^{0\nu}$ of the $0\nu2\beta$ decay of
$^{100}$Mo is one of the largest among 35 possible $2\beta^-$
decay candidates \cite{Doi85,Suh98}. Theoretical predictions for
the product of half-life with the effective neutrino mass
$T_{1/2}^{0\nu}\cdot\langle m_\nu\rangle^2$ are in the range of
$8.0\times10^{22}$ to $4.1\times 10^{24}$ yr$\cdot$eV$^2$
(see compilations \cite{Tre02} and more recent calculations
\cite{NMErecent})\footnote{One very deviating result is also known:
$3.6\times 10^{27}$ yr$\cdot$eV$^2$ in accordance with \cite{Pan96}.}.
Moreover, from the experimental point of view, the
larger is the $Q_{2\beta}$ energy, the simpler is to overcome
background problems, in particular, because background from
natural radioactivity drops sharply above 2615 keV, the energy of
$\gamma$'s from $^{208}$Tl decay ($^{232}$Th family). In addition,
cosmogenic activation, which is important for the next generation
$2\beta$ decay experiments (see, for instance \cite{CARVEL}),
contributes less at higher energies.

Two neutrino double beta decay ($2\nu 2\beta$) of $^{100}$Mo to
the ground state of $^{100}$Ru was already observed in a few
direct experiments \cite{100-gs} with measured half-lives in the
range of $3.3\times10^{18} - 1.2 \times10^{19}$ yr; the most exact
value comes from the recent measurements in the NEMO-3 experiment
as $T_{1/2}=7.1\pm0.5 \times10^{18}$ yr \cite{Arn05}. Geochemical
measurements gave the value of $T_{1/2}=2.2\pm0.3 \times10^{18}$
yr \cite{Hid04}. In addition to transition to the ground state,
also $2\nu 2\beta$ decay of $^{100}$Mo to the first excited
$0^+_1$ level of $^{100}$Ru ($E_{exc}=1131$ keV) was observed;
measured values of half-lives are in the range of $(5.7-9.3)
\times10^{20}$ yr \cite{100-exc}. The neutrinoless $2\beta$ decay
still is not observed: the highest limit was reached in the NEMO-3
experiment as $T_{1/2}>4.6 \times10^{23}$ yr at 90\% C.L.
\cite{Arn05}.

The purpose of our work was investigation of energy resolution,
light yield, $\alpha/\beta$ ratio, pulse shape for $\gamma$ rays
and $\alpha$ particles, temperature dependence of scintillation
properties, pulse-shape discrimination ability with a few samples
of CaMoO$_4$ crystal scintillators produced by the Institute for
Materials (IM, Lviv, Ukraine), and by the Innovation Centre of the
Moscow Steel and Alloy Institute (ICMSAI, Moscow, Russia).
Radioactive contamination of four samples of CaMoO$_4$ crystals
was tested in the Solotvina Underground Laboratory.

\section{SAMPLES}

The main properties of CaMoO$_4$ scintillators are presented in
Table 1, where characteristics of calcium and cadmium tungstates
are also given for comparison. The material is non-hygroscopic and
chemically resistant. Five clear, practically colorless CaMoO$_4$
crystals were used in our studies. The crystals were fabricated
from single crystals grown by the Czochralski method. All crystals
used in the present study are listed in Table 2.

\begin{table}[tbp]
\caption{Properties of CaMoO$_4$, CaWO$_4$, and CdWO$_4$ crystal
scintillators.}
\begin{center}
\begin{tabular}{|l|l|l|l|}
\hline ~                                              & CaMoO$_4$
& CaWO$_4$      & CdWO$_4$  \\ \hline Density (g/cm$^3$) & 4.2 &
6.1           & 8.0    \\ Melting point ($^\circ$C) & 1430       &
$1570-1650$     &  1325   \\ Structural type & Scheelite   &
Scheelite & Wolframite \\ Cleavage plane & Weak (001) & Weak (101)
& Marked (010) \\ Hardness (Mohs) & $3.5-4$    & $4.5-5$     &
$4-4.5$
\\ Wavelength of emission maximum (nm)            & 520        &
$420-440$   & 480     \\ Refractive index
& 1.98       & 1.94 & $2.2-2.3$  \\ Effective average decay
time$^{\ast}$ ($\mu$s) & 14         & 8           & 13  \\
Photoelectron yield [\% of NaI(Tl)]$^{\ast}$   & 9\%        & 18\%
& 20\%    \\ \hline \multicolumn{3}{l}{$^{\ast}$For $\gamma$ rays,
at room temperature.} \\
\end{tabular}
\end{center}
\end{table}

\begin{table}[tbp]
\caption{Samples of CaMoO$_4$ crystal scintillators used in this
study, and their scintillation properties.}
\begin{center}
\begin{tabular}{|l|l|l|l|l|l|}
\hline

 ID     & Size               & Mass & Manufac-   & Relative pulse   & FWHM at  \\

 ~      & ~(mm)              & (g)  & turer      & amplitude        & 662 keV  \\

\hline

CMO--1 & $25\times 13\times 9$    & 11.5 & IM$^a$     & 100\%          & 12.8\%$^c$\\
CMO--2 & $\oslash 38\times 20$    & 95.6 & IM$^a$     & 103\%          & 12.5\%$^d$ \\
CMO--3 & $\oslash 38\times 20$    & 99.9 & IM$^a$     & 110\%          & 11.9\%$^d$ \\
CMO--4 & $\oslash 38\times 20$    & 97.8 & IM$^a$     & 118\%          & 10.3\%$^d$ \\
CMO--5 & $28\times 28\times 24$   & 82.5 & ICMSAI$^b$ & 79\%           & 14.0\%$^c$ \\
\hline
\multicolumn{6}{l}{$^{a}$ Institute for Materials (Lviv, Ukraine)} \\
\multicolumn{6}{l}{$^{b}$ Innovation Centre of the Moscow
Steel and Alloy Institute (Moscow, Russia)} \\
\multicolumn{6}{l}{$^{c}$ exit surface (coupled to PMT) was polished; other surfaces were diffused} \\
\multicolumn{6}{l}{$^{d}$ exit and opposite surfaces were polished; side surface was diffused} \\
\end{tabular}
\end{center}
\end{table}

\section{MEASUREMENTS AND RESULTS}

\subsection{Energy resolution}

In the present work, the energy resolution was measured for all
the CaMoO$_4$ samples.

The CaMoO$_4$ crystal (CMO--4) was diffused at the side surface
with the help of fine grinding paper, the exit and top surfaces
were polished. The scintillator was wrapped by PTFE reflector tape
and optically coupled to 3" photomultiplier (PMT) Philips XP2412.
The measurements were carried out with the help of the home-made
spectroscopy amplifier with 16 $\mu$s shaping time to collect most
of the charge from the anode of the PMT. The scintillator was
irradiated by $\gamma$ quanta of $^{137}$Cs, $^{207}$Bi,
$^{232}$Th, and $^{241}$Am sources. The energy resolutions (full
width at half maximum, FWHM) of 34\% ($^{241}$Am, 60 keV), 10.3\%
($^{137}$Cs, 662 keV), 7.7\% ($^{207}$Bi, 1064 keV), and 4.7\%
($^{208}$Tl, 2615 keV) were measured (see Fig.~1). The energy
resolution obtained in the present study is the best ever reported
for CaMoO$_4$ crystal scintillators. It is important to stress
that the clear peak of 2615 keV $\gamma$ line of $^{208}$Tl
($^{232}$Th source) was accumulated with the energy resolution of
4.7\%. It demonstrates a possibility to calibrate the energy scale
of a CaMoO$_4$ detector in the vicinity of the expected peak of
$0\nu2\beta$ decay of $^{100}$Mo with the help of a $^{228}$Th
source.

\nopagebreak
\begin{figure}[t]
\begin{center}
\mbox{\epsfig{figure=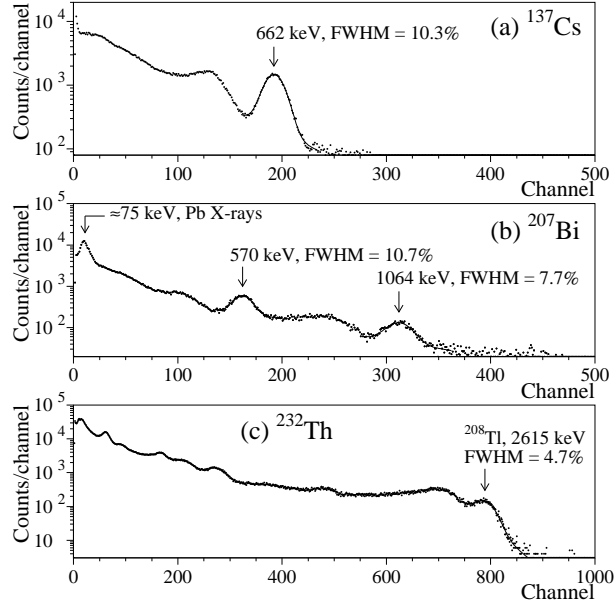,height=8.0cm}} \caption{Energy
spectra of $^{137}$Cs (a), $^{207}$Bi (b), and $^{232}$Th (c)
$\gamma$ quanta measured with CaMoO$_{4}$ scintillation crystal
$\oslash 38\times 20$ mm (CMO--4).}
\end{center}
\end{figure}

Dependence of energy resolution and light output on surface
treatment was checked with the sample CMO--5. First, all surfaces
of the crystal were polished. The crystal was wrapped by three
layers of PTFE tape and optically connected to PMT XP2412. The
spectroscopy amplifier with 16 $\mu$s shaping time was used. The
energy resolution of 16.2\% was measured in this conditions for
$^{137}$Cs 662 keV $\gamma$ line. Then surface of the crystal,
except of exit window connected to the PMT, was diffused by
fine-grained grinding paper. The relative pulse amplitude was
increased at 1.21 times and the energy resolution was improved to
14.0\%.

\subsection{\it Light yield}

The relative photoelectron yield of the CaMoO$_4$ crystal (CMO--4)
was measured relatively to the CaWO$_4$ scintillator
$\oslash40\times39$ mm described in \cite{Zdes05}. The crystals
were coupled to PMT XP2412 with the bialkali photocathode and were
irradiated by $\gamma$ quanta of $^{137}$Cs source. The
spectroscopy amplifier with the shaping time 16 $\mu$s was used in
this test. The relative photoelectron yield of the CaMoO$_4$
scintillator was measured as 36\% of the CaWO$_4$.

Photoelectron yield of the CMO--4 scintillator was also measured
relatively to the NaI(Tl) scintillator $\oslash $40$\times $40 mm
of standard assembling. To avoid an effect of the substantial
difference in the scintillation decay of CaMoO$_4$ ($14~\mu$s) and
NaI(Tl) (0.25 $\mu$s), the energy spectra were built by
calculating areas of pulses (over 300 $\mu$s for CaMoO$_4$ and 6
$\mu$s for NaI(Tl)) accumulated with the help of 20 MS/s transient
digitizer based on the 12 bit ADC (AD9022) \cite{Fazz98}. The
photoelectron yield of 8\% relatively to NaI(Tl) was obtained for
the CMO--4 sample.

\subsection{\it $\alpha /\beta$ ratio}

The $\alpha/\beta$ ratio was measured with the CMO--1 crystal by
using collimated $\alpha$ particles of a $^{241}$Am source. The
dimensions of the collimator were $\oslash $0.75 $\times $~2 mm.
As it was checked by a surface-barrier detector, the energy of
$\alpha$ particles was reduced to about 5.25 MeV by 2 mm of air
due to passing through the collimator \cite{Dane03}. Fig.~2 shows
the energy spectrum of the $\alpha$ particles measured by the
CaMoO$_4$ scintillator. The $\alpha/\beta$ ratio is 0.20 at the
energy of $\alpha$ particles 5.25 MeV.

\nopagebreak
\begin{figure}[t]
\begin{center}
\mbox{\epsfig{figure=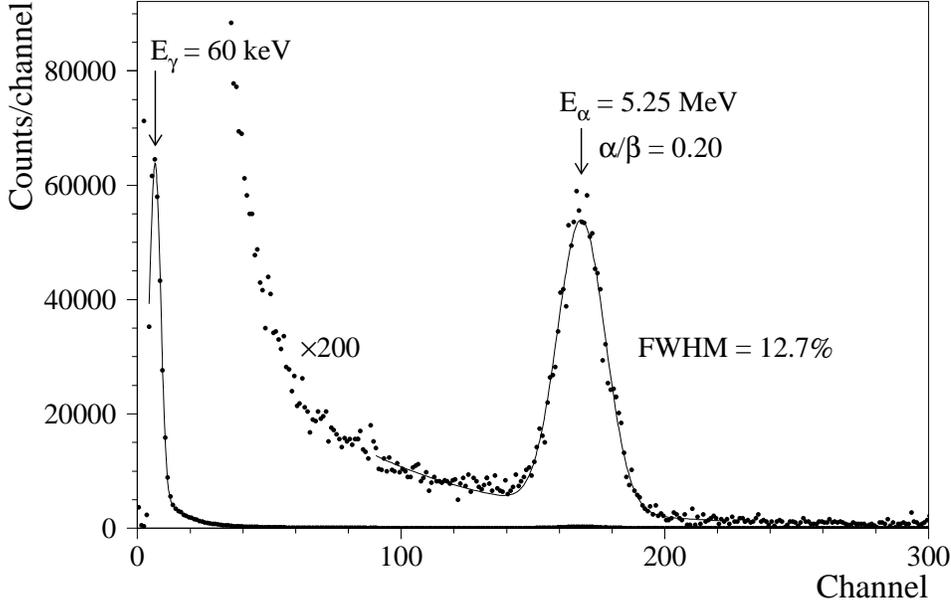,height=8.0cm}} \caption{The energy
spectrum measured with 5.25 MeV $\alpha$ particles from $^{241}$Am
source (dots). Fits of $\alpha$ and 60 keV $\gamma$ peaks are
shown by solid lines.}
\end{center}
\end{figure}

Besides the measurements with the external source, $\alpha $ peaks
of $^{220}$Rn and $^{216}$Po (from $^{232}$Th chain), $^{210}$Po
and $^{214}$Po ($^{238}$U), presented in trace amount in the
CaMoO$_4$ crystal ($\oslash 38\times 20$ mm, CMO--2), were used to
extend the energy range of $\alpha $ particles. The peaks of
$^{220}$Rn, $^{216}$Po and $^{214}$Po were selected with the help
of the time-amplitude analysis of data obtained in the low
background measurements (see subsection 3.7.3). The clear peak of
$^{210}$Po is presented in the energy spectrum of CaMoO$_4$
detector (subsection 3.7.2). The measured dependence of the
$\alpha/\beta$ ratio on energy of $\alpha$ particles (Fig.~3) can
be described in the energy region 5--8 MeV by the linear function:
$\alpha/\beta=0.11(2) + 0.019(3)E_{\alpha}$, where $E_{\alpha}$ is
energy of $\alpha$ particles in MeV.

\nopagebreak
\begin{figure}[t]
\begin{center}
\mbox{\epsfig{figure=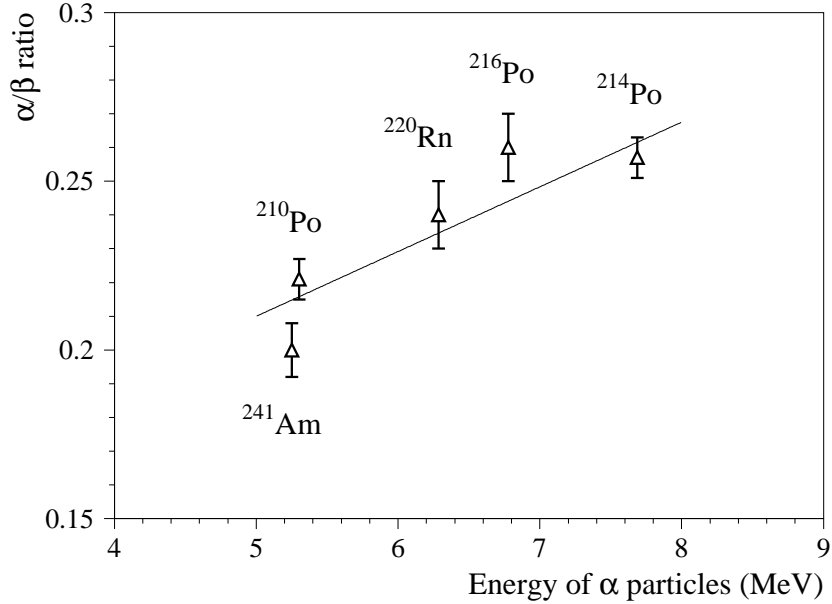,height=8.0cm}} \caption{Dependence
of $\alpha/\beta$ ratio for CaMoO$_4$ scintillator on energy of
$\alpha$ particles.}
\end{center}
\end{figure}

\subsection{\it Pulse shape for $\gamma$ rays and $\alpha$
particles}

Pulse shape of CaMoO$_4$ scintillator (CMO--1) was studied with
the help of the 12 bit transient digitizer operated at 20 MS/s. To
study pulse shape of scintillation decay for $\alpha$ particles,
the CaMoO$_4$ crystal was irradiated by $\alpha$ particles from
collimated $^{241}$Am source. A $^{60}$Co source was used to
investigate pulse shape for $\gamma$ quanta. Measurements were
carried out at the temperature $(27\pm1)~^\circ$C.

\nopagebreak
\begin{figure}[t]
\begin{center}
\mbox{\epsfig{figure=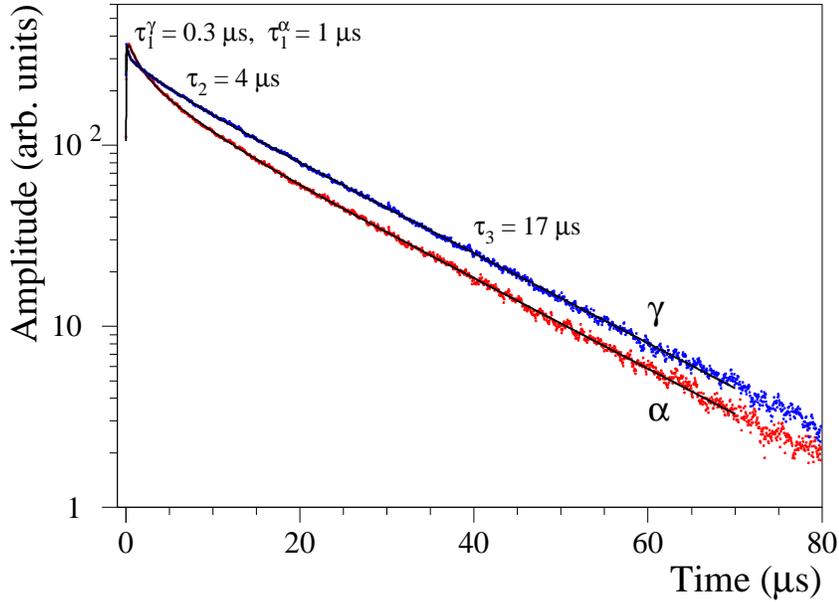,height=8.0cm}} \caption{Decay of
scintillation in CaMoO$_4$ crystal for $\gamma$ rays and $\alpha$
particles measured by 20 MS/s transient digitizer. Three
components of scintillation signals with time decay of 0.3--1
$\mu$s, 4 $\mu$s and 17 $\mu$s are shown. Fitting functions for
$\alpha$ and $\gamma$ pulses are shown by solid lines.}
\end{center}
\end{figure}

The shape of the light pulses produced by $\alpha$ particles and
$\gamma$ rays in the CaMoO$_4$ scintillator measured by the 20
MS/s digitizer are shown in Fig.~4. To obtain the pulse shapes,
about one thousand of individual $\alpha$ ($\gamma$) events with
amplitudes corresponding to $\alpha$ peak of $^{241}$Am were
summed. The pulses were fitted by the function:
\begin{center}
$f(t)=\sum A_{i}(e^{-t/\tau _{i}}-e^{-t/\tau
_{0}})/(\tau_{i}-\tau_{0}),\qquad t>0$,
\end{center}
where $A_{i}$ are the relative intensities, $\tau_{i}$ are the
decay constants for different light-emission components, and
$\tau_{0}$ is integration constant of electronics
($\tau_{0}\approx~0.08~\mu$s). Three decay components were
observed with $\tau_{i}\approx 0.3-1~\mu$s, $\approx4~\mu$s, and
$\approx17~\mu$s with different intensities for $\gamma$ rays and
$\alpha$ particles (see Fig.~4 and Table~3).

\begin{table}[tbp]
\caption{Decay time of CaMoO$_4$ scintillators for $\gamma$ quanta
and $\alpha$ particles measured by 20 MS/s transient digitizer at
the temperature $+27~^\circ$C. The decay constants and their
relative intensities are denoted as $\tau_i$ and A$_i$,
respectively.}
\begin{center}
\begin{tabular}{|l|l|l|l|}
\hline
Type of irradiation & \multicolumn{3}{|c|}{Decay constants and relative intensities}  \\
\cline{2-4}
~                   & $\tau_1$~(A$_1$)     & $\tau_2$~(A$_2$)   & $\tau_3$~(A$_3$) \\
\hline
$\alpha$ particles  & $1~\mu$s ($2\%$)     & $4~\mu$s ($13\%$)  & $17~\mu$s ($85\%$)  \\
$\gamma$ rays       & $0.3~\mu$s ($0.5\%$) & $4~\mu$s ($5.5\%$) & $17~\mu$s ($94\%$)  \\
\hline
\end{tabular}
\end{center}
\end{table}

\subsection{\it Pulse-shape discrimination between $\gamma$ rays
and $\alpha$ particles}

The difference of the pulse shapes allows to discriminate $\gamma
$($\beta$) events from those induced by $\alpha$ particles. We
applied for this purpose the optimal filter method proposed in
\cite{Gatt62}, and successfully used for different scintillation
detectors: CdWO$_4$ \cite{Fazz98,Bard06_CWO}, CeF$_3$
\cite{Bel03}, CaWO$_4$ \cite{Zdes05}, YAG:Nd \cite{Dane05_YAG},
ZnWO$_4$ \cite{ZWO}, CaF$_2$(Eu) \cite{Bell06}, PbWO$_4$
\cite{Bard06_PWO}. For each CaMoO$_4 $ signal, a numerical
characteristic (shape indicator, $SI$) was calculated in the
following way:

\begin{center}
$SI=\sum f(t_k) P(t_k)/\sum f(t_k)$,
\end{center}
where the sum is over time channels $k,$ starting from the origin
of pulse and up to 50 $\mu$s, $f(t_k)$ is the digitized amplitude
(at the time $t_k$) of the signal. The weight function $P(t)$ was
defined as: $P(t)=\{f_{\alpha}(t)-f_{\gamma}(t)\}/\{f_{\alpha}
(t)+f_{\gamma} (t)\}$, where $f_{\alpha} (t)$ and $f_{\gamma} (t)$
are the reference pulse shapes for $\alpha$ particles and $\gamma
$ quanta.

Reasonable discrimination between $\alpha$ particles and $\gamma$
rays was achieved using this approach, as one can see in Fig.~5
where the $SI$ distributions measured with the CaMoO$_4$
scintillation crystal CMO--1 for $\alpha$ particles
($E_{\alpha}\approx 5.25$ MeV) and $\gamma$ quanta ($\approx 1$
MeV) are shown. As a measure of discrimination ability (factor of
merit, $FOM$), the following expression can be used:

\nopagebreak
\begin{figure}[t]
\begin{center}
\mbox{\epsfig{figure=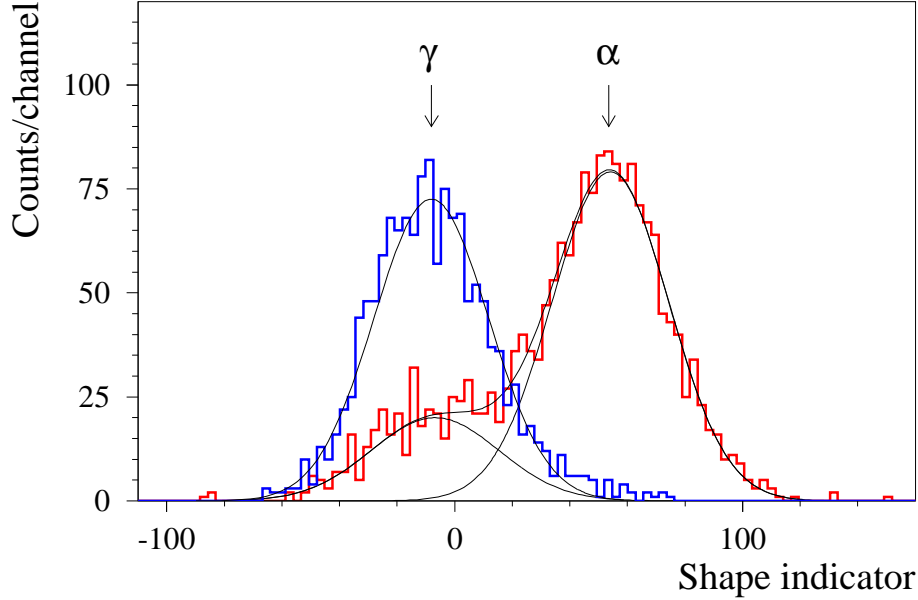,height=8.0cm}} \caption{The shape
indicator (see text) distributions measured by CaMoO$_4$ detector
with $\alpha$ particles ($E_{\alpha}=5.25$ MeV) and $\gamma$
quanta ($\approx 1$ MeV) using 20 MS/s 12 bit transient digitizer.
The distributions were fitted by Gaussian function (solid lines).
The small tail in the shape indicator distribution for $\alpha$ at
$\approx-10$ can be explained by background $\gamma$ events.}

\end{center}
\end{figure}

\begin{center}
 $FOM=\mid SI_{\alpha}-SI_{\gamma}\mid/\sqrt{\sigma_{\alpha}^2+\sigma_{\gamma}^2}$,
\end{center}
where $SI_{\alpha}$ and $SI_{\gamma}$ are mean $SI$ values for
$\alpha$ particles and $\gamma$ quanta distributions (which are
well described by Gaussian functions), $\sigma_{\alpha}$ and
$\sigma_{\gamma}$ are the corresponding standard deviations. For
the distributions presented in Fig.~5, the factor of merit
$FOM=2.2$.

\subsection{Temperature dependence of light output and pulse shape}

Temperature dependence of scintillation properties were studied in
the range of $-158~\div~+25~^\circ$C with the CMO--1 sample. The
crystal was viewed by PMT XP2412 through a high purity quartz
light-guide 4.9 cm in diameter and 25 cm long. The CaMoO$_4$
scintillator and the light-guide were wrapped by PTFE tape as
reflector. Dow Corning Q2-3067 optical couplant was used to
provide optical contact of the scintillator with the light-guide,
and of the light-guide with the PMT. The crystal and main part of
the light-guide were placed into a Dewar vessel, while PMT was
isolated from the vessel with the help of a foam plastic plate. In
such a way the temperature of PMT was kept stable (practically at
room temperature) during the measurements. The Dewar vessel was
periodically filled by small portions of liquid nitrogen to cool
the detector. The temperature of the crystal was measured with the
help of a chromel-alumel thermocouple.

To measure the relative light output and pulse shape, the
scintillator was irradiated by $\alpha$ particles of the
collimated $^{241}$Am source ($E_{\alpha}=5.25$ MeV). The 20 MS/s
transient digitizer was used to accumulate four thousands
scintillation pulse shapes of CaMoO$_4$ at each temperature. Then
the recorded pulse shapes were used to build energy spectra and to
determine the decay time. The energy spectra were obtained by
calculation of the area of each signal from the starting point up
to 380 $\mu$s.

To present behaviour of scintillation decay with temperature, we
use the averaged decay time $\langle \tau \rangle$ determined by
the following formula:

\begin{center}
 $\langle \tau \rangle=\sum (\tau_i A_i)/\sum A_i$.
\end{center}

The obtained dependence of averaged decay time on temperature is
presented in Fig.~6 (a), and can be fitted in the temperature
interval $-127~\div~+25~^\circ$C by the function $\langle \tau
\rangle=31.8(7)-0.85(3)\times T+0.0093(7)\times T^2$, where $T$ is
temperature in degree C. At the temperature $-50~^\circ$C (where
the maximum of light output was observed) the averaged decay time
is $\approx98~\mu$s. The dependence of the averaged decay time on
temperature for CaMoO$_4$ scintillator agrees qualitatively with
the result reported in Ref. \cite{Belo05}.

The measured dependence of relative light output of CaMoO$_4$
crystal scintillator on temperature is presented in Fig.~6 (b).
The relative light output ($RLO$) increases with decreasing of the
temperature down to $\approx -50~^\circ$C as
$RLO=1.374(16)-0.0137(8)\times T-0.00005(3)\times T^2$, where $T$
is temperature in degrees C. Then some decrease of scintillation
intensity was observed. The dependence is in agreement with the
result obtained in Ref. \cite{Belo05}.

\nopagebreak
\begin{figure}[t]
\begin{center}
\mbox{\epsfig{figure=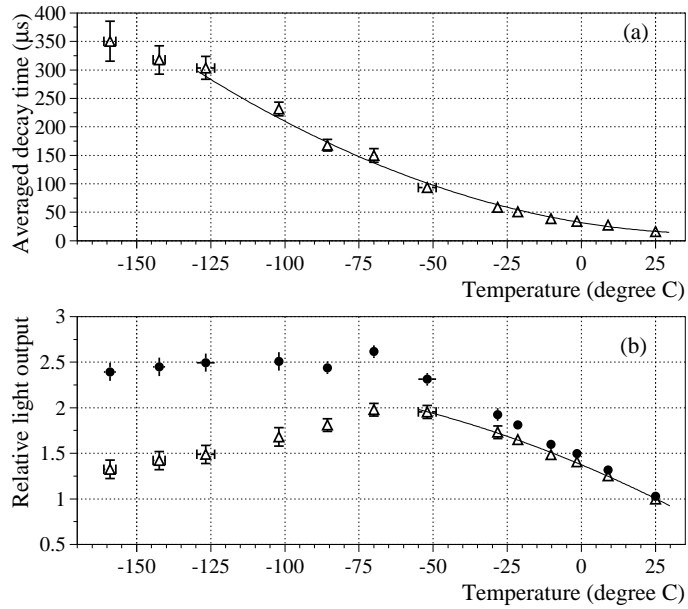,height=8.0cm}} \caption{(a)
Temperature dependence of averaged decay time measured with
CaMoO$_4$ detector under irradiation by $\alpha$ particles of
$^{241}$Am source. (b) The dependence of relative light output on
temperature (triangles). The data corrected taking into account
decay time of scintillation signals are shown by filled circles.
Solid lines represent fits of the data.}
\end{center}
\end{figure}

Temperature dependence of the radioluminescence intensity was also
studied with the help of photon counting method. The dependence
was measured in the temperature range $-170~\div~+40~^\circ$C
under gamma-excitation by $^{57}$Co source. Light from sample was
directed to the PMT (FEU-100) input window through a condenser.
The $^{57}$Co source was installed at 8 cm distance from the
sample. PMT counting rate was averaged during 10 s at each value
of the temperature. The dependence of the luminescence yield is
presented in Fig. 7. It was found that at temperature decrease
down to $-80~^\circ$C the light yield increases by the factor of
2.5. Light yield remains practically unchanged at the further
temperature decrease to $-170~^\circ$C.

\nopagebreak
\begin{figure}[t]
\begin{center}
\mbox{\epsfig{figure=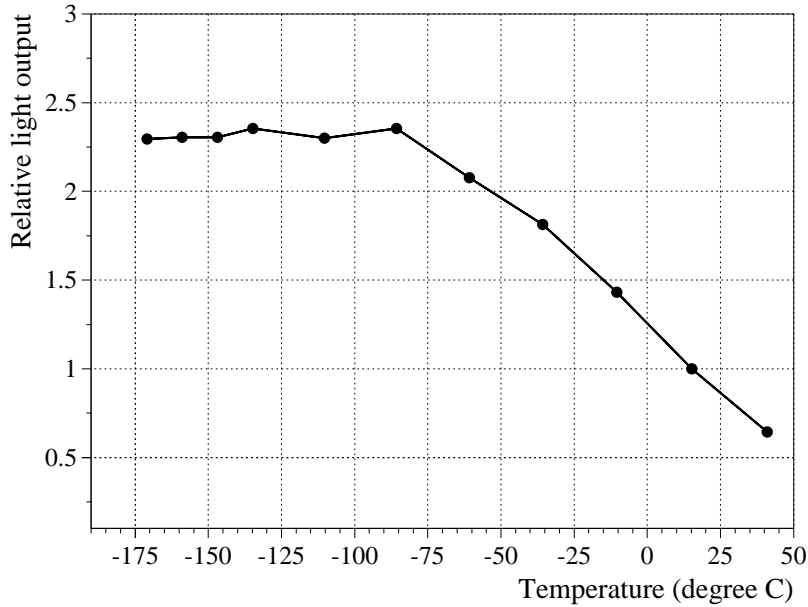,height=8.0cm}} \caption{Temperature
dependence of radioluminescence intensity of CaMoO$_4$ crystal
measured with the help of photon counting method.}
\end{center}
\end{figure}

The difference in the temperature dependence of light yield
measured by two methods can be explained by considerable increase
of the scintillation decay time at temperatures lower than
$\sim-50~^{\circ}$C. Whereas result of the photon counting method
(with time window 10 s) does not depend on the decay time, the
relative light output obtained by using the transient digitizer
(when scintillation pulses were integrated over 380 $\mu$s)
depends on the kinetics of scintillation decay. After this effect
was taken into account, the behaviour of the relative pulse
amplitude measured with the digitizer (the corrected data are
presented in Fig. 6 (b) by filled circles) are practically the
same as measured with the help of the photon counting method.

\subsection{Radioactive contamination}

\subsubsection{Set-up and measurements}

Radiopurity of four samples (CMO--2, CMO--3, CMO--4, and CMO--5)
was tested in the Solotvina Underground Laboratory built in a salt
mine 430 m underground ($\simeq $1000 m of water equivalent)
\cite{Zde87}.

The crystals CMO--2, CMO--3, and CMO--4 were produced from one
boule with aim to study a possible dependence of radioactive
contamination on length of the boule. Such a dependence was
observed for CdWO$_4$ crystals produced for the Solotvina
experiment to search for $2\beta$ decay of $^{116}$Cd
\cite{Dan95}. The crystal CMO--2 was cut from the top of the boule
(beginning of growth), the CMO--3 was taken from the middle part,
and the CMO--4 sample was produced from the bottom part of the
monocrystal boule.

Radioactive contamination of the CaMoO$_4$ crystals was measured
in the low background set-up installed in the Solotvina
Underground Laboratory. In the set-up a scintillation CaMoO$_4$
crystal was viewed by the special low-radioactive 5$^{^{\prime
\prime }}$ photomultiplier tube (EMI D724KFLB) through the high
purity quartz light-guide $\oslash 10\times 33$ cm. The detector
was surrounded by a passive shield made of teflon (thickness of
3--5 cm), plexiglass (6--13 cm), high purity (OFHC) copper (3--6
cm), lead (15 cm) and polyethylene (8 cm). For each event occurred
in the detector, the amplitude of a signal and arrival time were
recorded. In addition, CaMoO$_4$ scintillation pulse shapes were
digitized with a 20 MHz sampling frequency in $\approx 100~\mu$s
full range. Energy scale of the detectors was calibrated with the
help of $^{207}$Bi $\gamma $ source through the calibration
channel made in the shield. For instance, the energy resolution of
the CMO--2 crystal was FWHM = 17.8\% and 13.5\% for 570 and 1064
keV $\gamma$ lines, respectively.

\nopagebreak
\begin{figure}[t]
\begin{center}
\mbox{\epsfig{figure=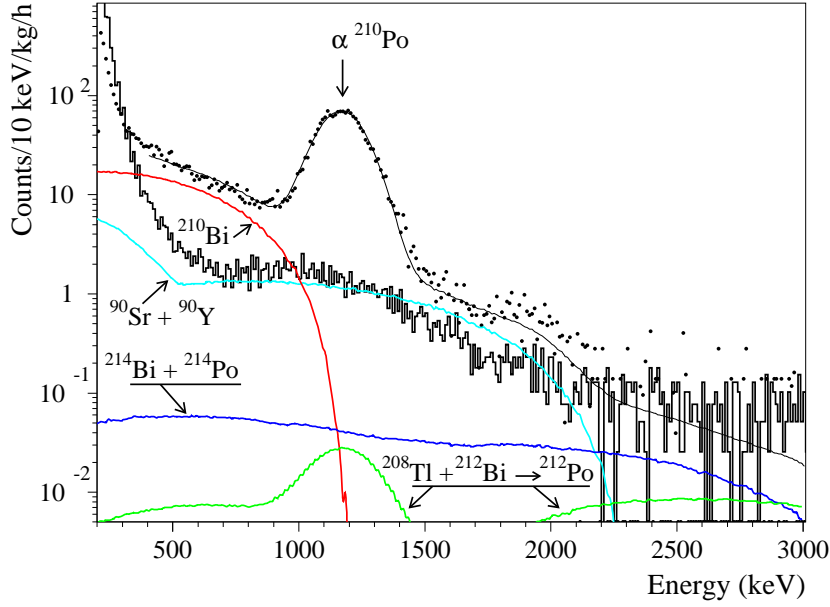,height=8.0cm}} \caption{Energy
spectrum of CaMoO$_4$ scintillation crystal (CMO--2, dots)
measured in the low background set-up during 74.83 h. Solid line
represents the fit of the data by the background model in the
$0.4-3$ MeV energy interval. The peak with the energy
$\approx1.17$ MeV is mainly due to the $\alpha$ decay of
$^{210}$Po. Most important components of the background
($^{210}$Bi, $^{90}$Sr$+^{90}$Y, $^{208}$Tl +
$^{212}$Bi$\rightarrow ^{212}$Po, and $^{214}$Bi + $^{214}$Po) are
shown. The spectrum measured during 472.4 h with the crystal
CMO--5 in the same conditions is drown by solid histogram. Both
spectra are normalized on mass of the crystals and time of
measurements.}
\end{center}
\end{figure}

\subsubsection{Interpretation of background spectrum}

The energy spectrum of the CMO--2 detector measured during 74.83 h
in the low background set-up is presented in Fig.~8. The peak at
the energy $\approx 1.17$ MeV can be attributed to intrinsic
$^{210}$Po (daughter of $^{210}$Pb from the $^{238}$U family) with
activity of 0.42(1) Bq/kg. The equilibrium of the uranium chain in
the crystal was broken during the crystal production, because
there is no peak of $^{238}$U expected at the energy of $\approx
0.82$ MeV (in the gamma scale). Analysis of the spectrum gives
only limit for activity of $^{238}$U on the level of $\leq0.5$
mBq/kg. The same situation is with peaks of the uranium's
daughters $^{234}$U, $^{230}$Th, and $^{226}$Ra, which cannot be
resolved (their $Q_\alpha $ values are very close). A common for
these nuclides $\alpha$ peak is expected at the energy $\approx1$
MeV. Fit of the spectrum gives again only the limit for the total
activity of these isotopes at the level of 2.8 mBq/kg. In the same
way the following limits on the activities of $^{222}$Rn,
$^{218}$Po, and $^{232}$Th were obtained: $\leq4.4$ mBq/kg,
$\leq4.2$ mBq/kg, and $\leq0.7$ mBq/kg, respectively.

To take into account presence in the crystal of $\beta$ active
isotopes (from U/Th families, $^{40}$K, $^{90}$Sr$+^{90}$Y), the
energy spectrum of the CaMoO$_4$ detector was simulated with the
GEANT4 package \cite{GEANT4}. Initial kinematics of particles
emitted in $\beta$ decays of nuclei was generated with the DECAY0
event generator \cite{DECAY0}. The spectrum of the CMO--2 crystal
(Fig.~8) was fitted in the energy interval $0.4-3$ MeV by the
model, which includes the simulated distributions of U/Th
daughters ($^{208}$Tl and $^{212}$Bi$\rightarrow ^{212}$Po,
$^{210}$Bi, $^{214}$Bi + $^{214}$Po, $^{234m}$Pa), $^{40}$K,
$^{90}$Sr$+^{90}$Y, Gaussian function to describe the $\alpha$
peak of $^{210}$Po, and an exponential function to take into
account external $\gamma$ background. The main components of the
background are shown in Fig.~8. The major part of the $\beta $
activity can be ascribed to $^{210}$Bi, daughter of $^{210}$Pb
($\approx $0.4 Bq/kg). We cannot also exclude presence of
significant activity of $^{90}$Sr$+^{90}$Y in the crystal.
Nevertheless, because there are no clear peculiarities which can
be used to prove presence of these nuclides, we can give only the
limit on activities of $^{90}$Sr$+^{90}$Y and $^{210}$Bi in the
crystal at the level of $\leq 62$ mBq/kg and $\leq 398$ mBq/kg,
respectively.

\subsubsection{Time-amplitude analysis}

The raw background data were analyzed by the time-amplitude
method, when the energy and arrival time of each event were used
for selection of some decay chains in $^{232}$Th and $^{238}$U
families\footnote{Technique of a time-amplitude analysis of
background data to recognize a presence of the short-living chains
from $^{232}$Th, $^{235}$U and $^{238}$U families was described in
\cite{Dan95,GSO}.}. For instance, the following sequence of
$\alpha $ decays from the $^{232}$Th family was searched for and
observed: $^{220}$Rn ($Q_\alpha $ = $6.41$ MeV, $T_{1/2}$ = $55.6$
s) $ \rightarrow $ $^{216}$Po ($Q_\alpha $ = $6.91$ MeV, $T_{1/2}$
= $0.145$ s) $\rightarrow $ $^{212}$Pb. These radionuclides are in
equilibrium with $^{228}$Th from $^{232}$Th family. Because the
energy of $\alpha $ particles from $^{220}$Rn decay corresponds to
$\simeq $1.5 MeV in $\gamma $ scale of the CaMoO$_4$ detector, the
events in the energy region 1.4 -- 2.2 MeV were used as triggers.
Then all events (within 1.4 -- 2.2 MeV) following the triggers in
the time interval 0.02 -- 0.6 s (containing 84\% of $^{216}$Po
decays) were selected. The obtained $\alpha $ peaks (see Fig. 9)
are in agreement with those expected for $\alpha $ particles of
$^{220}$Rn $\rightarrow $ $^{216}$Po $\rightarrow $ $^{212}$Pb
chain \cite{ToI98}. The pulse-shape analysis confirms the nature
of the events as caused by $\alpha$ particles. On this basis, in
spite of low statistic, the activity of $^{228}$Th in the
CaMoO$_4$ crystal can be calculated as 0.23(10) mBq/kg.

\nopagebreak
\begin{figure}[t]
\begin{center}
\mbox{\epsfig{figure=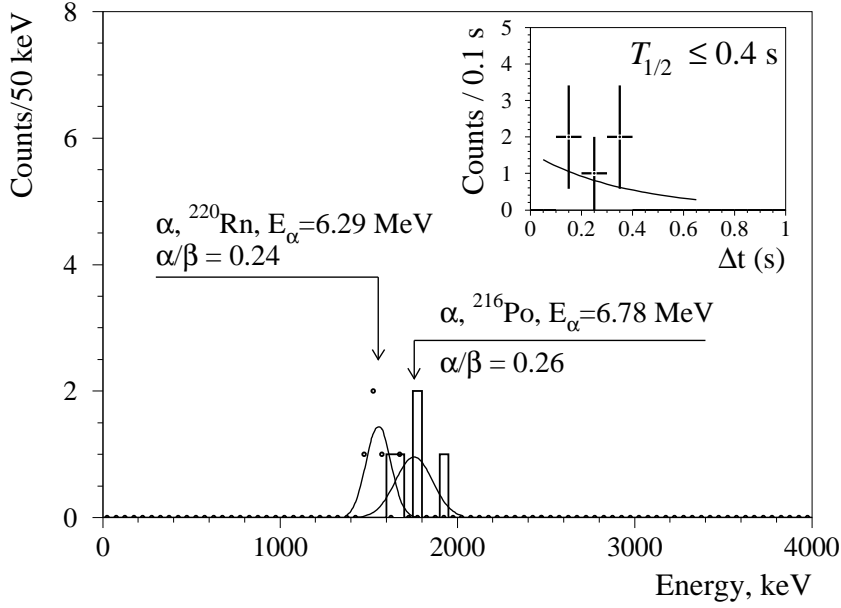,height=8.0cm}} \caption{The energy
distributions for the fast sequence of the $\alpha$ decays of
$^{220}$Rn and $^{216}$Po selected from the background data by the
time-amplitude analysis. (Inset) The time distribution between the
first and second events together with an exponential fit.}
\end{center}
\end{figure}

Similarly, for the analysis of the $^{226}$Ra chain ($^{238}$U
family) the following sequence of $\beta $ and $\alpha $ decays
was used: $^{214}$Bi ($Q_\beta =3.27$ MeV) $\rightarrow $
$^{214}$Po ($Q_\alpha =7.83$ MeV, $ T_{1/2}=164$ $\mu $s)
$\rightarrow $ $^{210}$Pb. For the first event the lower energy
threshold was set at 0.25 MeV, while for the events of the
$\alpha$ decay of $^{214}$Po the energy window $1.4-4$ MeV was
chosen. The events were selected in the time interval of $100-800$
$\mu$s (55\% of $^{214}$Po decays). Taking into account the
registration efficiency for events of $^{214}$Bi with the energy
threshold 0.25 MeV (85\%), the obtained spectra (Fig.~10) lead to
the $^{226}$Ra activity in the CaMoO$_4$ crystal of 2.1(4) mBq/kg.

\nopagebreak
\begin{figure}[t]
\begin{center}
\mbox{\epsfig{figure=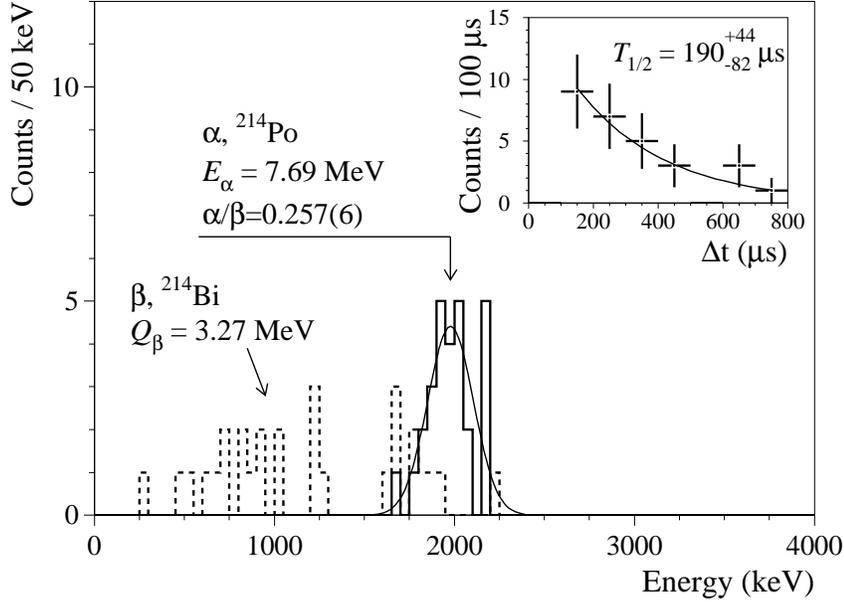,height=8.0cm}} \caption{The energy
distributions for the fast sequence of the $\beta$ ($^{214}$Bi)
and $\alpha$ ($^{214}$Po) decays selected from the background data
by the time-amplitude analysis. (Inset) The time distribution
between the first and second events together with an exponential
fit. The obtained half-life of $^{214}$Po ($190^{+44}_{-82}~\mu$s)
is in agreement with the table value ($164~\mu$s) \cite{ToI98}.}
\end{center}
\end{figure}

The results obtained with the help of time-amplitude analysis
don't contradict the results of the analysis described in
subsection 3.7.2 if one takes into account broken equilibrium of
$^{232}$Th and $^{238}$U chains.

Estimation of radioactive contamination in other crystals was done
in the same way as described above. The summary of the measured
radioactive contamination of the CaMoO$_4$ scintillators (or
limits on their activities) is given in Table~4, again in
comparison with CaWO$_4$ \cite{Zdes05} and CdWO$_4$ detectors
\cite {Geo96,Bur96,Dan96,Dan03}.

\begin{table}[htbp]
\caption{Radioactive contaminations in CaMoO$_4$, CaWO$_4$, and
CdWO$_4$ crystals scintillators.}
\begin{center}
\begin{tabular}{|l|l|l|l|l|l|l|}
\hline
Source      & \multicolumn{6}{|c|}{Activity (mBq/kg)}     \\
\cline{2-7}
~           & CMO-2       & CMO--3     & CMO--4       & CMO--5     & CaWO$$ \cite{Zdes05} &  CdWO$$ \cite{Geo96,Bur96,Dan96,Dan03}  \\
\hline
~           & ~           & ~          & ~            &   ~        &   ~                  &   ~        \\
$^{232}$Th: & ~           & ~          & ~            &   ~        &   ~                  &   ~        \\
$^{232}$Th  & $\leq0.7$   & $\leq0.7$  &  $\leq0.9$   & $\leq 1.5$ & 0.69(10)             & 0.053(5)   \\
$^{228}$Th  & 0.23(10)    & 0.42(17)   & 0.4(4)       & 0.04(2)    & 0.6(2)               & $\leq 0.004-0.039(2)$ \\
~           & ~           & ~          & ~            &   ~        &   ~                  &   ~        \\
$^{238}$U:  & ~           & ~          & ~            &   ~        &   ~                  &   ~        \\
$^{238}$U   &  $\leq 0.5$ & $\leq 0.6$ &   $\leq 0.6$ & $\leq 1.5$ & 5.6(5)               & $\leq 0.004$ \\
$^{226}$Ra  &  2.1(4)     &  2.5(5)    &  2.4(1.3)    &  0.13(4)   & 5.6(5)               & $\leq 0.004$ \\
$^{210}$Pb  & $\leq 398$  & $\leq 401$ & $\leq 550$   & $\leq17$   & $\leq 430$           & $\leq 0.4$ \\
$^{210}$Po  &  420(10)    & 490(10)    & 550(20)      & $\leq8$    & 291(5)               & $\leq 0.4$ \\
~           & ~           & ~          & ~            &   ~        &   ~                  &   ~        \\
$^{40}$K    & $\leq 1.1 $ & $\leq 2.1$ & $\leq 2.5$   & $\leq 3$   & $\leq 12$            & 0.3(1)     \\
$^{90}$Sr   & $\leq 62  $ & $\leq 178$ & $\leq 50$    & $\leq 23$  & $\leq 70$            & $\leq 0.2$ \\
~           & ~           & ~          &  ~           &   ~        &   ~                  &  ~         \\
\hline
\end{tabular}
\end{center}
\end{table}

One can see that radioactive impurities in the CaMoO$_4$ crystals
are comparable with CaWO$_4$ crystals, and are much higher (by
factor of 10 -- 10$^3$) than those of the CdWO$_4$ scintillators.
The radioactive contamination of CaMoO$_4$ crystal produced in the
Innovation Centre of Moscow Steel and Alloy Institute is lower
than that of the scintillators produced in the Institute for
Materials. Some indication on increasing of radioactive
contamination in the crystal volume during the growth process was
observed with the samples CMO--2, CMO--3, and CMO--4 produced from
the same crystal boule.

\section{High sensitivity $^{100}$Mo $0\nu 2\beta$ experiment with CaMoO$_4$ detectors}

Below we discuss possible use of CaMoO$_4$ scintillators as potential
detectors in search for neutrinoless 2$\beta$ decay of $^{100}$Mo.
Because of the "source=detector" approach (which provides high efficiency
for detection of the process), good energy resolution and pulse-shape discrimination
ability (which allows to reduce background), CaMoO$_4$ scintillators
could be considered as a promising tool in a $^{100}$Mo $0\nu 2\beta$ experiment.
To estimate sensitivity of the experiment,
in the following we present calculations of some backgrounds from cosmogenic
activities induced in CaMoO$_4$ crystals, as well as from internal pollution by the
U/Th chains, and from two neutrino 2$\beta$ decays of $^{100}$Mo and $^{48}$Ca.
Natural composition of Ca and O, and 100\% enrichment in $^{100}$Mo
is supposed.

Finite energy resolution of CaMoO$_4$ detector not only causes a broadening of
$^{100}$Mo $0\nu 2\beta$ peak, but also results in presence of events from tail of
$2\nu 2\beta$ distribution in the peak's region. This background is unavoidable
because it is caused by $^{100}$Mo itself; it could be minimized only by improvement
of the energy resolution of the detector. The response functions of CaMoO$_4$
scintillator for
two neutrino and neutrinoless 2$\beta$ decays of $^{100}$Mo are presented in
Fig.~11 for 3\%, 4\%, 5\% and 6\% energy resolution (FWHM) of the detector
at the energy of $^{100}$Mo $0\nu 2\beta$ decay. Amplitude of $2\nu 2\beta$
($0\nu 2\beta$) distribution corresponds to $T_{1/2}(2\nu)=7\times10^{18}$ yr
\cite{Arn05} ($T_{1/2}(0\nu)=1\times10^{24}$ yr).
It is evident from Fig.~11 that energy resolution should not be greater than 4--5\%.
For FWHM=4\%, number of events from $2\nu 2\beta$ tail in 1~FWHM energy interval
centered at $Q_{2\beta}$ is equal to 0.06 events per 1~kg of Ca$^{100}$MoO$_4$
crystal per 1~year.

Contribution from $2\nu 2\beta$ decay of $^{48}$Ca could be much
more dangerous (see Fig.~12). Measured half-life of this process
is equal $T_{1/2}(2\nu)=4\times10^{19}$ yr (see f.e. compilation
\cite{Tre02}); $^{48}$Ca is present in natural Ca composition with
abundance of 0.187\% \cite{Boh05}. Event's rate from $^{48}$Ca
$2\nu 2\beta$ decay in 1~FWHM energy interval is equal to 1.4
events per 1~kg of Ca$^{100}$MoO$_4$ crystal per 1~year.

\nopagebreak
\begin{figure}[ht]
\begin{center}
\mbox{\epsfig{figure=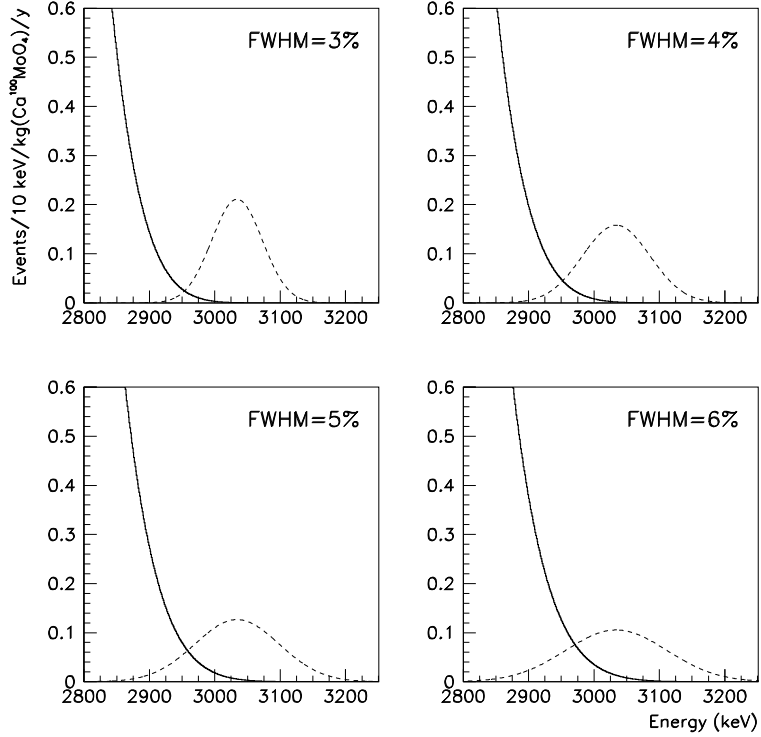,width=10.0cm}}
\caption{The response functions of a Ca$^{100}$MoO$_4$ detector for
2$\beta$ decays of $^{100}$Mo for
$T_{1/2}(2\nu)=7\times10^{18}$ yr (solid lines) and
$T_{1/2}(0\nu)=1\times10^{24}$ yr (dashed lines) for different energy resolutions
of the detector at the energy of $^{100}$Mo $0\nu 2\beta$ decay.}
\end{center}
\end{figure}

In estimations of backgrounds from internal U/Th chains and cosmogenic
activities, we additionally suppose use of an active shield made of
CsI(Tl) scintillators. Radiopure CsI(Tl) scintillators were developed by
the KIMS collaboration for the dark matter experiments in the Yangyang Laboratory
\cite{KIMS}. In simulations, the CaMoO$_4$ crystal
$\oslash$45$\times$45 mm was placed in the center of
$\oslash$40$\times$40 cm CsI(Tl) scintillation detector. CaMoO$_4$ was
viewed by two PMTs through light-guides also made of CsI(Tl)\footnote{Another
possible solution could be light-guides made of PbWO$_4$ scintillators
as it was proposed in \cite{Dane06}.}.
Such an active shield suppresses internal and external backgrounds related with
emission of $\gamma$ quanta. Contribution from cosmic ray particles in course of
measurements could be effectively suppressed by deep underground location of the
experiment.

Only two isotopes in U/Th chains have enough energy to create background in the
region of $0\nu 2\beta$ peak of $^{100}$Mo: $^{208}$Tl ($Q_\beta$=5001 keV
\cite{ToI98}) and $^{214}$Bi ($Q_\beta$=3272 keV)\footnote{Formally also
$^{210}$Tl is present with $Q_\beta$=5489 keV but yield of this isotope
in $^{238}$U chain is only 0.021\%.}. Beta decay of $^{208}$Tl is accompanied
by emission of one or more $\gamma$ quanta, and thus it is effectively
eliminated by the CsI(Tl) active shield. Beta decay of $^{214}$Bi is more dangerous
because it has 18.2\% branch to the ground state of $^{214}$Po without emission
of any $\gamma$'s and with $Q_\beta$=3272 keV. However, this contribution could
be further suppressed by $\simeq$1 order of magnitude by checking for fast
$\alpha$ decay of $^{214}$Po ($T_{1/2}=164.3$ $\mu$s) during subsequent $\simeq$1 ms and
proving its $\alpha$ nature with the pulse-shape analysis. Contributions
from $^{208}$Tl and $^{214}$Bi are shown in Fig.~12 for 0.1 mBq/kg activity.
It is clear that they are not very dangerous in comparison with
$^{48}$Ca $2\nu 2\beta$ decay because comparative or even more clean
CaMoO$_4$ crystals were already obtained
(CMO-5 in Table~4: 0.04 mBq/kg for $^{208}$Tl and 0.13 mBq/kg for $^{214}$Bi).

\nopagebreak
\begin{figure}[ht]
\begin{center}
\mbox{\epsfig{figure=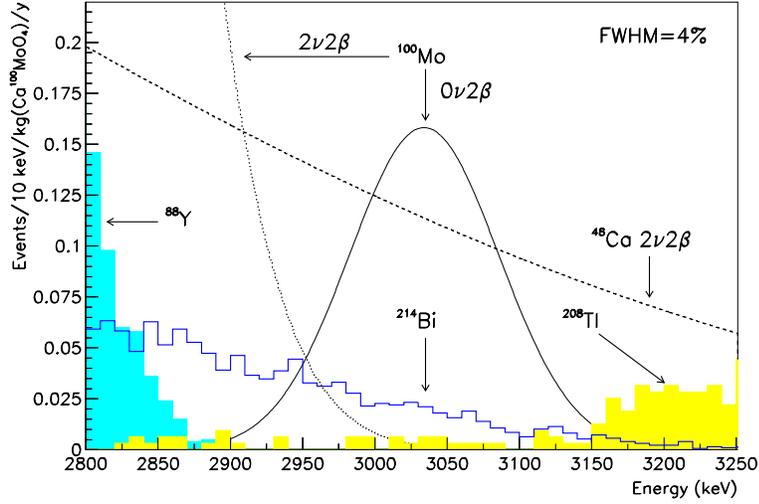,width=10.0cm}}
\caption{Calculated backgrounds from
$2\nu 2\beta$ decay of $^{48}$Ca ($T_{1/2}(2\nu)=4\times10^{19}$ yr),
internal pollutions by $^{208}$Tl and $^{214}$Bi (both with 0.1 mBq/kg),
and $^{88}$Y isotope from cosmogenic activity.
One order of magnitude suppression for $^{214}$Bi with
the pulse-shape analysis is taken into account.
Amplitude of $^{88}$Y distribution corresponds to 1000 decays in CaMoO$_4$,
much bigger than that expected from the $^{88}$Y cosmogenic activity
(4 events per year on average during the first 5 years).
It is supposed that Ca$^{100}$MoO$_4$ scintillator is operating in
anticoincidence with the CsI(Tl) active shield (see text).
Distributions for $^{100}$Mo are shown for
$T_{1/2}(2\nu)=7\times10^{18}$ yr and
$T_{1/2}(0\nu)=1\times10^{24}$ yr.}
\end{center}
\end{figure}

Cosmogenic activities produced by cosmic rays in Ca$^{100}$MoO$_4$ crystal
during its time production period at the Earth surface were calculated with
the COSMO code \cite{Mar92}. An activation time of 30 days at sea level, and
a deactivation time of 1 year underground were assumed. The most dangerous
cosmogenic nuclides -- with energy close or higher than $Q_{2\beta}$ of
$^{100}$Mo and noticeable yield -- are summarized in Table~5. Simulations
with GEANT4 \cite{GEANT4} and initial kinematics given by the DECAY0
event generator \cite{DECAY0} showed that contributions from cosmogenic
activities are small in comparison with $2\nu 2\beta$ decay of $^{48}$Ca
(see Fig.~12 for $^{88}$Y).

To estimate sensitivity of the experiment to $^{100}$Mo $0\nu
2\beta$ decay in terms of half-life limit, we can use known
formula: $\lim T_{1/2} = \ln 2 \cdot \eta \cdot N \cdot t/\lim S$,
where $\eta$ is the detection efficiency, $N$ is the number of
$^{100}$Mo nuclei, $t$ is the measuring time, and $\lim S$ is the
maximum number of $0\nu 2\beta$ events which can be excluded with
a given confidence level on the basis of the experimental data or
simulated background. It is interesting to consider here only
unremovable backgrounds from $2\nu 2\beta$ decays of $^{48}$Ca and
$^{100}$Mo itself, neglecting all other external and internal
backgrounds (which could be effectively suppressed by the active
CsI(Tl) shield).

\begin{table}[t]
\caption{Cosmogenic radioactivity induced in Ca$^{100}$MoO$_4$
crystals. Exposition during 30 days to cosmic rays at the sea
level and 1 yr period of cooling down in underground conditions
are supposed. $D_5$ is number of decays during the first 5 years
of data taking per 1 kg of the crystal.}
\begin{center}
\begin{tabular}{|lllll|}
\hline
Isotope or               & $T_{1/2}$ & \multicolumn{2}{l}{Decay mode}    & $D_5$ \\
chain of decays          &           & \multicolumn{2}{l}{and $Q$ value} &       \\
\hline
$^{22}$Na                & 2.6 yr    & EC        & 2842 keV              & 8.99  \\
$^{42}$Ar\&$^{42}$K      & 32.9 yr   & $\beta^-$ & 3525 keV              & 0.80  \\
$^{56}$Co                & 77.3 d    & EC        & 4565 keV              & 0.01  \\
$^{60}$Co                & 5.3 yr    & $\beta^-$ & 2824 keV              & 2.08  \\
$^{68}$Ge\&$^{68}$Ga     & 270.8 d   & EC        & 2921 keV              & 0.90  \\
$^{88}$Y                 & 106.7 d   & EC        & 3623 keV              & 20.69 \\
\hline
\end{tabular}
\end{center}
\end{table}

We will have 1.41 (0.06) events from $2\nu 2\beta$ decay of $^{48}$Ca ($^{100}$Mo)
in 1 kg Ca$^{100}$MoO$_4$ scintillator inside 1~FWHM interval centered at the $^{100}$Mo
$Q_{2\beta}$ energy during 1 year.
In case of absence of other contributions, with measured
1 (or 2) events and expected background of 1.47 events, in accordance with the Feldman-Cousins
procedure \cite{Fel98}, value of $\lim S$ is equal to 2.9 (4.4) at 90\% C.L. Taking into account
that this interval contains 0.761 of the full $^{100}$Mo $0\nu 2\beta$ peak, it gives
half-life limit as: $T_{1/2}(0\nu)>5.4\times10^{23}$ ($3.5\times10^{23}$) yr.
Thus, comparatively modest efforts with 1 kg crystal (which contains near 490 g of $^{100}$Mo)
and 1 yr measurement time could give an interesting $T_{1/2}$ limit
which can be compared with the recent value from the NEMO-3
experiment: $4.6\times10^{23}$ yr \cite{Arn05} obtained with 7 kg of $^{100}$Mo after 389 d of
data taking.

Final aim of the NEMO-3 experiment ($T_{1/2}(0\nu)>2\times10^{24}$
yr) could be achieved with Ca$^{100}$MoO$_4$ scintillator at
statistics of 10~kg$\cdot$yr, i.e. also in a middle-scale
experiment. However, further improvement will be difficult task:
half-life limit of $10^{25}$ yr could be reached only at~200
kg$\cdot$yr statistics. More sensitive searches for $^{100}$Mo
$0\nu 2\beta$ decay evidently will need in depletion of Ca in
$^{48}$Ca\footnote{Only 20 years ago, before the first laboratory
observation of $2\nu 2\beta$ decay in 1987, it was difficult to
imagine that this rarest observed in nature process could be a
serious background in searches for more rare decays.}.

CaMoO$_4$ crystals could be also used as scintillating bolometers
\cite{Pirr06}. In this case energy resolution will be much better
($\simeq$5 keV instead of $\simeq$120 keV for 4\%) that results in
more clear interpretation of backgrounds and higher sensitivity of
an experiment. Two neutrino $2\beta$ decay of $^{100}$Mo will not
contribute anymore to the $0\nu 2\beta$ peak. With background only
from $2\nu 2\beta$ decay from non-depleted $^{48}$Ca, the
following half-life limits could be reached (at 90\% C.L.):
$6.5\times10^{23}$ yr, $4.1\times10^{24}$ yr, and
$4.2\times10^{25}$ yr for 1, 10 and 200 kg$\cdot$yr statistics,
respectively. Statistics of 1000 kg$\cdot$yr would correspond to
$T_{1/2}(0\nu)>1.1\times10^{26}$ yr. In accordance with the NME
calculations for $^{100}$Mo $0\nu 2\beta$ decay
\cite{Tre02,NMErecent}, it will give limit on the effective
neutrino mass in the range of 0.03--0.20 eV.

\section{CONCLUSIONS}

Scintillation properties and radioactive contamination of
CaMoO$_4$ crystals produced by the Institute for Materials (Lviv,
Ukraine), and by the Innovation Centre of Moscow Steel and Alloy
Institute (Moscow, Russia) were studied.

The energy resolution 10.3\%, 7.7\%, and 4.7\% for the 662, 1064,
and 2615 keV $\gamma $ lines was obtained with the CaMoO$_4$
sample of $\oslash38\times20$ mm produced by the Institute for
Materials. To our knowledge, such an energy resolution was never
reported for CaMoO$_4$ crystal scintillators.

The photoelectron yield of CaMoO$_4$ scintillator at room
temperature was measured as 36\% of CaWO$_4$ crystal, and
$\approx8\%$ of NaI(Tl).

The $\alpha/\beta$ ratio was measured with $\alpha$ particles of
$^{241}$Am $\alpha$ source, and by using $\alpha$ peaks from
internal U/Th contamination. The dependence of the $\alpha/\beta$
ratio on energy of particles ($E_{\alpha}$) in the energy interval
5--8 MeV can be described by the linear function
$\alpha/\beta=0.11(2)+0.019(3)E_{\alpha}$.

Three components of scintillation decay ($\tau_{i}\approx 0.3-1$,
$\approx4$ and $\approx17~\mu$s) and their intensities under
$\alpha$ particles and $\gamma$ quanta irradiation were measured
by using transient digitizer with 20 MHz sampling frequency. It
allows to discriminate $\alpha$ particles and $\gamma$ quanta with
reasonable efficiency.

The temperature dependence of light output and pulse shape were
measured in the range of temperature $-158~\div~+25~^\circ$C by
recording the pulse shapes with subsequent off-line analysis. At
the temperature of approximately $-50~^{\circ}$C the light output
increases in $\approx2$ times comparatively to light output at
room temperature. Averaged decay time increases with decreasing of
temperature as $\langle \tau
\rangle=31.8(7)-0.85(3)T+0.0093(7)\times T^2$, where $T$ is
temperature in degree C.  At the temperature $-50~^{\circ}$C
(where the maximum of light output was observed) the averaged
decay time for $\alpha$ particles of 5.25 MeV is $\approx98~\mu$s.
Temperature dependence of radioluminescence intensity of CaMoO$_4$
crystal was also investigated in the range of temperatures
$-170~\div~+40~^\circ$C with the help of photon counting method by
averaging PMT counting rate over 10 s. It was found that light
yield increases in $\approx2.5$ times at temperature
$-80~^\circ$C, and then remains practically unchanged at the
further temperature decrease to $-170~^\circ$C. We interpret such
a difference in the behaviour of light yield on temperature
measured by two methods as a result of very slow scintillation
decay at temperatures lower than $\sim-50~^{\circ}$C.

Radioactive contamination of CaMoO$_4$ crystals was estimated in
low background measurements carried out in the Solotvina
Underground Laboratory. CaMoO$_4$ scintillators produced in the
Institute for Materials (Lviv, Ukraine) are considerably polluted
by uranium and thorium (particularly by $^{210}$Po at the level of
$\approx 0.4-0.5$ Bq/kg). The contamination of CaMoO$_4$ crystal
produced by the Innovation Centre of the Moscow Steel and Alloy
Institute (Moscow, Russia) is one--two order of magnitude better.
It was found that equilibrium in uranium chains is broken in
CaMoO$_4$ crystals. Some indication on increasing of radioactive
contamination in the crystal volume during the growth process
was observed.

Perspectives for a high sensitivity experiment to search for
$0\nu2\beta$ decay of $^{100}$Mo are discussed. The energy
resolution of 4--5\% is enough to reach sensitivity at the level
of $10^{25}$ yr. The contamination of crystals by $^{226}$Ra and
$^{232}$Th should not exceed the level of 0.1 mBq/kg.
Two neutrino mode of $2\beta$ decay of $^{48}$Ca
restricts sensitivity of an experiment to search for $0\nu2\beta$
decay of $^{100}$Mo with the help of CaMoO$_4$ crystal
scintillators. A possible solution would be production of
CaMoO$_4$ scintillators from Calcium depleted in $^{48}$Ca.
Further improvement of sensitivity could be reached by applying CaMoO$_4$
crystals as scintillating bolometers.

A R\&D with aim to develop low-background, high resolution
scintillation detector for an experiment to search for
$0\nu2\beta$ decay of $^{100}$Mo with a sensitivity at the level
of $10^{25}$ yr is in progress.

\section{ACKNOWLEDGMENT}

Work of A.N.~Annenkov, O.A.~Buzanov, V.N.~Kornoukhov, M.~Korzhik
and O.~Missevitch was supported in part by ISTC project \#3293 in
collaboration with the Dark Matter Research Center and School of
Physics of Seoul National University, Republic of Korea.


\begin{thebibliography}{199}

\bibitem{Dan89}        F.A.~Danevich et al., Instr. Exp. R. 32 (1989) 1059.
\bibitem{Bur95}        S.F.~Burachas et al., Phys. Atom. Nucl. 58 (1995) 153.
\bibitem{Dan96}        F.A.~Danevich et al., Z. Phys. A 355 (1996) 433.
\bibitem{Bel99}        P.~Belli et al., Nucl. Phys. B 563 (1999) 97.
\bibitem{GSO}          F.A.~Danevich et al., Nucl. Phys. A 694 (2001) 375.
\bibitem{Dan03}        F.A.~Danevich et al., Phys. Rev. C 68 (2003) 035501.
\bibitem{Bel03}        P.~Belli et al., Nucl. Instr. Meth. A 498 (2003) 352.
\bibitem{Oga04}        I.~Ogawa et al., Nucl. Phys. A 730 (2004) 215.
\bibitem{ZWO}          F.A.~Danevich et al., Nucl. Instr. Meth. A 544 (2005) 553.
\bibitem{Belo05}       S.~Belogurov et al., IEEE Nucl. Sci. 52 (2005)
1131;\\
                       H.J. Kim, et al., Proceedings of New View in Particle Physics (VIETNAM2004), August 5-11, 2004, p.
                       449.
\bibitem{Mikh05-pssb}  V.B.~Mikhailik et al., Phys. Stat. Sol. b 242 (2005) R17.
\bibitem{Mikh05-JAP}   V.B.~Mikhailik et al., J. Appl. Phys. 97 (2005) 083523.
\bibitem{Mikh05-JPCM}  V.B.~Mikhailik et al., J. Phys.: Condens. Matter 17 (2005) 7209.
\bibitem{Seny06}       A.~Senyshyn et al., Phys. Rev. B 73 (2006) 014104.
\bibitem{Mikh06-JPDAP} V.B.~Mikhailik and H.~Kraus, J. Phys. D: Appl. Phys. 39 (2006) 1181.
\bibitem{Pirr06}       S.~Pirro et al., Phys. Atom. Nucl. 69 (2006) 2109.
\bibitem{Doi85}        M.~Doi et al., Prog. Theor. Phys. Suppl. 83 (1985) 1.
\bibitem{Suh98}        J.~Suhonen and O.~Civitarese, Phys. Rep. 300 (1998) 123.
\bibitem{Tre02}        V.I.~Tretyak and Yu.G.~Zdesenko, At. Data Nucl. Data Tables 61 (1995) 43; 80 (2002) 83.
\bibitem{NMErecent}    F.~Simkovic et al., Phys. Rev. C 64 (2001) 035501;\\
                       S. Stoica and V.P. Paun, Rom. Journ. Phys. 47 (2002) 497;\\
                       J.~Suhonen, Nucl. Phys. A 700 (2002) 649;\\
                       O.~Civitarese and J.~Suhonen, Nucl. Phys. A 729 (2003) 867;\\
                       V.A. Rodin et al., Phys. Rev. C 68 (2003) 044302;\\
                       V.A. Rodin et al., Nucl. Phys. A 766 (2006) 107; erratum arXiv: 0706.4304v1 [nucl-th].
\bibitem{Pan96}        G. Pantis et al., Phys. Rev. C 53 (1996) 695.
\bibitem{CARVEL}       Yu.G.~Zdesenko et al., Astropart. Phys. 23 (2005) 249.
\bibitem{100-gs}       S.I.~Vasil'ev et al., JETP Lett. 51 (1990) 622; \\
                       H.~Ejiri et al., Phys. Lett. B 258 (1991) 17; \\
                       D.~Dassie et al., Phys. Rev. D 51 (1995) 2090; \\
                       M.~Alston-Garnjost et al., Phys. Rev. C 55 (1997) 474; \\
                       A.~De Silva et al., Phys. Rev. C 56 (1997) 2451; \\
                       V.D. Ashitkov et al., JETP Letters 74 (2001) 529.
\bibitem{Arn05}        R. Arnold et al., Phys. Rev. Lett. 95 (2005) 182302.
\bibitem{Hid04}        H. Hidaka, C.V. Ly, K. Suzuki, Phys. Rev. C 70 (2004) 025501.
\bibitem{100-exc}      A.S.~Barabash et al., Phys. Lett. B 345 (1995) 408; \\
                       A.S.~Barabash et al., Phys. At. Nucl. 62 (1999) 2039; \\
                       M.J. Hornish et al., Phys. Rev. C 74 (2006) 044314; \\
                       R. Arnold et al. (NEMO Collaboration), Nucl. Phys. A 781 (2006) 209.
\bibitem{Zdes05}       Yu.G.~Zdesenko et al.,  Nucl. Instrum. Meth. A 538 (2005) 657.
\bibitem{Fazz98}       T.~Fazzini et al., Nucl. Instrum. Meth. A 410 (1998) 213.
\bibitem{Dane03}       F.A.~Danevich et al.,  Phys. Rev. C 67 (2003) 014310.
\bibitem{Gatt62}       E. Gatti and F. De Martini, Nuclear Electronics 2, IAEA, Vienna, 1962, p.~265.
\bibitem{Bard06_CWO}   L.~Bardelli et al.,  Nucl. Instr. Meth. A 569 (2006) 742.
\bibitem{Dane05_YAG}   F.A.~Danevich et al.,  Nucl. Instr. Meth. A 541 (2005) 583.
\bibitem{Bell06}       P.~Belli et al., Nucl. Phys. A 789 (2007) 15.
\bibitem{Bard06_PWO}   L.~Bardelli et al., arXiv:nucl-ex[0706.2422].
\bibitem{Zde87}        Yu.G.~Zdesenko et al., Proc. 2nd Int. Symp. Underground
                       Physics, Baksan Valley, USSR, August 17--19, 1987. -- Moscow,
                       Nauka, 1988, p. 291.
\bibitem{Dan95}        F.A.~Danevich et al., Phys. Lett. B 344 (1995) 72.
\bibitem{GEANT4}       S.~Agostinelli et al., Nucl. Instr. Meth. A 506 (2003) 250; \\
                       J. Allison et al., IEEE Trans. Nucl. Sci. 53 (2006) 270.
\bibitem{DECAY0}       O.A.~Ponkratenko, V.I.~Tretyak, Yu.G.~Zdesenko, Phys. At. Nucl. 63 (2000) 1282;
                       V.I.~Tretyak, to be published.
\bibitem{ToI98}        R.B. Firestone et al., {\it Table of Isotopes}, 8th ed.,
                       John Wiley \& Sons, New York, 1996 and CD update, 1998.
\bibitem{Geo96}        A.Sh.~Georgadze et al., Instrum. Exp. Tech. 39 (1996) 191.
\bibitem{Bur96}        S.Ph.~Burachas et al., Nucl. Instr. Meth. A 369 (1996) 164.
\bibitem{Boh05}        J.K. Bohlke et al., J. Phys. Chem. Ref. Data 34 (2005) 57.
\bibitem{KIMS}         H.S.~Lee et al., Phys. Lett. B 633 (2006)
201;\\
                       H.S.~Lee et al., Nucl. Instr. Meth. A 571 (2007) 644.
\bibitem{Dane06}       F.A.~Danevich et al., Nucl. Instr. Meth. A 556 (2006) 259.
\bibitem{Mar92}        C.J. Martoff and P.D. Lewin, Comp. Phys. Comm. 72 (1992) 96.
\bibitem{Fel98}        G.J.~Feldman and R.D.~Cousins, Phys. Rev. D 57 (1998) 3873.

\end{thebibliography}
\end{document}